\documentclass[3p]{elsarticle}

\usepackage{amsmath}
\usepackage{amsthm}
\usepackage{amssymb}
\usepackage{graphicx}
\usepackage{esint}
\usepackage{color}

\usepackage{lineno}
\PassOptionsToPackage{hyphens}{url}
\usepackage[pdfencoding=auto,psdextra,colorlinks=true]{hyperref}
\usepackage{bookmark}

\biboptions{numbers,sort&compress}

\journal{Brain Research}

\biboptions{authoryear}

\begin{document}

\begin{frontmatter}

\title{Computational capacity of pyramidal neurons in the cerebral cortex}

\author[address1]{Danko D. Georgiev\corref{mycorrespondingauthor}}
\ead{danko.georgiev@mail.bg}
\cortext[mycorrespondingauthor]{Corresponding author}

\author[address2]{Stefan K. Kolev}
\ead{kolevskk@abv.bg}

\author[address3]{Eliahu Cohen}
\ead{eliahu.cohen@biu.ac.il}

\author[address4]{James F. Glazebrook}
\ead{jfglazebrook@eiu.edu}

\address[address1]{Institute for Advanced Study, 30 Vasilaki Papadopulu Str., Varna 9010, Bulgaria}
\address[address2]{Institute of Electronics, Bulgarian Academy of Sciences,
72 Tzarigradsko Chaussee Blvd., Sofia 1784, Bulgaria}
\address[address3]{Faculty of Engineering and the Institute of Nanotechnology
and Advanced Materials, Bar Ilan University, Ramat Gan 5290002, Israel}
\address[address4]{Department of Mathematics and Computer Science, Eastern Illinois University, Charleston, IL 61920, USA}

\begin{abstract}
The electric activities of cortical pyramidal neurons are supported by structurally stable, morphologically complex axo-dendritic trees. Anatomical differences between axons and dendrites in regard to their length or caliber reflect the underlying functional specializations, for input or output of neural information, respectively. For a proper assessment of the computational capacity of pyramidal neurons, we have analyzed an extensive dataset of three-dimensional digital reconstructions from the NeuroMorpho.Org database, and quantified basic dendritic or axonal morphometric measures in different regions and layers of the mouse, rat or human cerebral cortex. Physical estimates of the total number and type of ions involved in neuronal electric spiking based on the obtained morphometric data, combined with energetics of neurotransmitter release and signaling fueled by glucose consumed by the active brain, support highly efficient cerebral computation performed at the thermodynamically allowed Landauer limit for implementation of irreversible logical operations. Individual proton tunneling events in voltage-sensing S4 protein $\alpha$-helices of Na$^{+}$, K$^{+}$ or Ca$^{2+}$ ion channels are ideally suited to serve as single Landauer elementary logical operations that are then amplified by selective ionic currents traversing the open channel pores. This miniaturization of computational gating allows the execution of over 1.2 zetta logical operations per second in the human cerebral cortex without combusting the brain by the released heat.
\end{abstract}


\date{July 26, 2020}

\end{frontmatter}

%\linenumbers

\section{Introduction}

The cerebral cortex is the seat of higher cognitive functions in mammals.
Structurally, it is divided into neocortex, made up of six layers
of neurons, and allocortex, made up of just three or four layers of
neurons \citep{Rockland2018}. The neocortex forms the largest, outer
layer of the cerebrum. In large mammals and primates, the neocortex
is folded into grooves and ridges, which minimize the brain volume,
and are pivotal for the wiring of the brain and its functional organization
\citep{Rakic2009}. The neocortex is involved in sensory perception,
awareness, attention, motor control, working memory, thought, intelligence,
and consciousness \citep{Page1981}. The allocortex includes evolutionary
older regions, such as the olfactory system and the hippocampus, which
comprise the neural basis of emotion and play important roles in time
ordering of memorized events or the consolidation of conscious memory
from short-term to long-term memory \citep{Fournier2015,Wible2013,Squire2015}.

Excitatory, glutamatergic pyramidal neurons are the principal type
of cell comprising over 70\% of all cortical neurons \citep{Nieuwenhuys1994}.
Pyramidal neurons, referred to as the ``psychic cells'' of the brain
by Ram\'{o}n y Cajal \citep{Goldman-Rakic2002}, are organized in complex
neuronal networks, which communicate by means of electric signals.
Wiring of the corresponding neuronal networks requires individual neurons to
support structurally stable, elongated cable-like projections referred
to as \emph{neurites}. Depending on their functional specialization, the
neurites could be classified as dendrites, specialized in delivering
inputs to the neuron, or axons, specialized in delivering outputs
from the neuron to other neurons \citep{Georgiev2017}. Dendrites deliver
electric signals through activated synapses mainly formed onto spines
of the dendritic tree \citep{Eyal2018}. The post-synaptic electric
currents propagate passively along the dendrites through an electrotonic
mechanism that summates the electric signals spatially and temporally
at the cell body (soma) of the neuron. Axons output electric spikes
(action potentials) in an active fashion that consumes large amounts
of biochemical energy in order to propagate the electric signals without attenuation at
a distance to pre-synaptic axonal buttons whose release of neurotransmitter
subsequently affects the electric properties of dendrites of target
neurons.

The morphology of neurites is intimately related to their characteristic
functional role \citep{Mounier2015}. Dendrites achieve processing of received information
through passive and lossy transmission. Consequently, the dendrites
have shorter lengths and larger diameters in order to compensate for
the electrotonic attenuation of currents with distance. Alternatively,
axons are required to deliver output signals at large distances to
target neurons through lossless transmission achieved at the expense
of biochemical energy. To reduce energy expenditure, axons are thinner
and insulated with myelin sheets. Thus, a detailed study of neuronal
morphology is essential for better understanding of the neuronal hardware
behind higher cognitive functions.

Here, we analyze a dataset of 749 three-dimensional neuronal reconstructions
from NeuroMorpho.org~7.8 digital archive \citep{Ascoli2007}.
Then, with the use of morphometric, electrophysiological and biochemical
data, we derive an upper bound on the computational capacity of pyramidal
neurons in the cerebral cortex. Finally, we conclude with a theoretical
discussion on the fundamental limitations imposed by energetics on
possible subneuronal mechanisms for the processing of cognitive information.

\begin{figure}[t!]
\begin{centering}
\includegraphics[width=165mm]{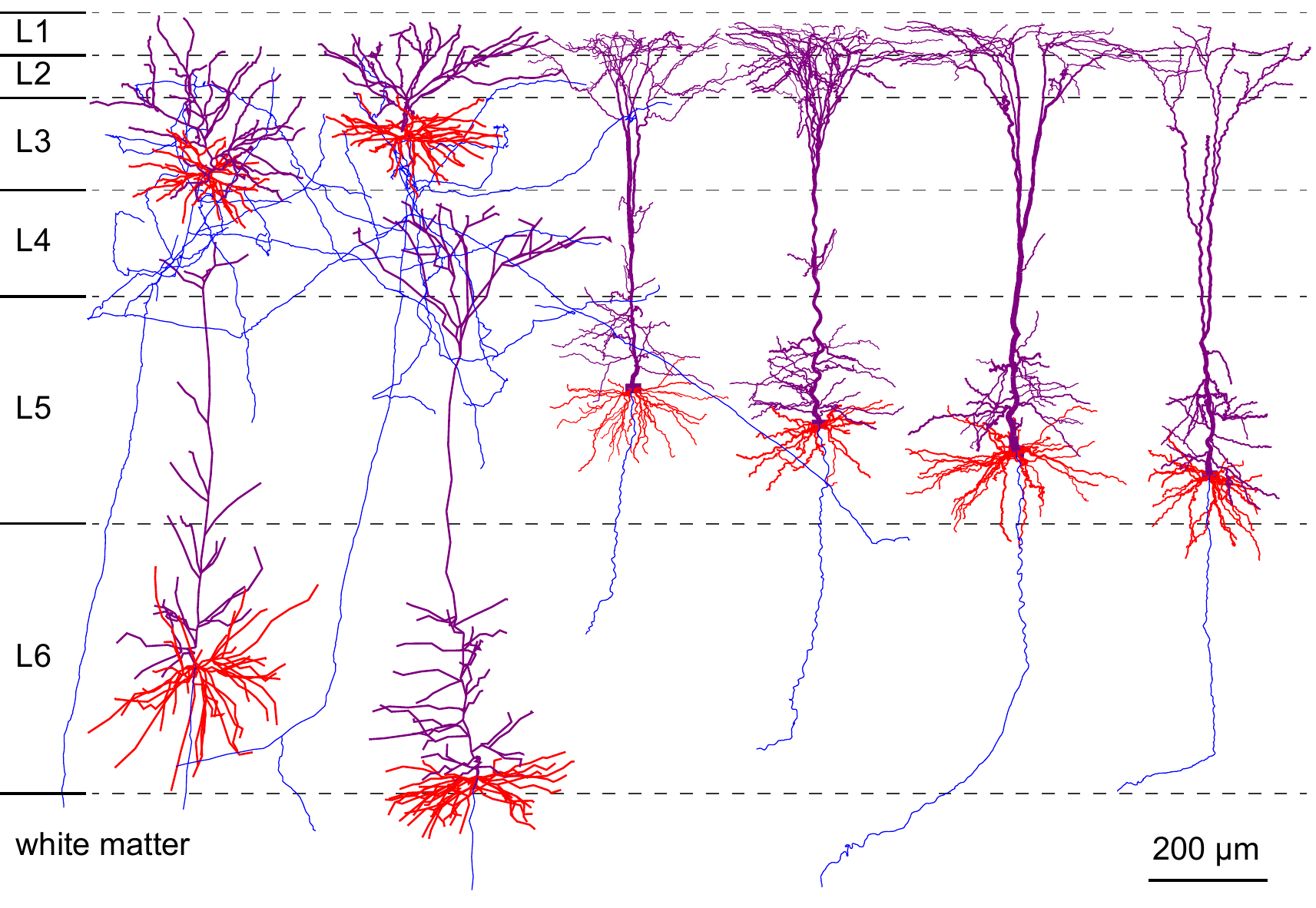}
\par\end{centering}

\caption{\label{fig:1}Layered structure of mouse neocortex constructed in
silico with digital reconstructions of Layer 2-3 pyramidal neurons
(NMO\_51117, NMO\_51116), Layer 5 pyramidal neurons (NMO\_09483, NMO\_09485,
NMO\_09480, NMO\_09494) and Layer 6 pyramidal neurons (NMO\_85158,
NMO\_85162). Basal dendrites are rendered in red, apical dendrites
in purple, and axons in blue. Neuron identification numbers are given
from left to right of the rendered reconstructions.}
\end{figure}

\section{Results}

\subsection{Dendrite morphometry}

\begin{figure}[t!]
\begin{centering}
\includegraphics[width=165mm]{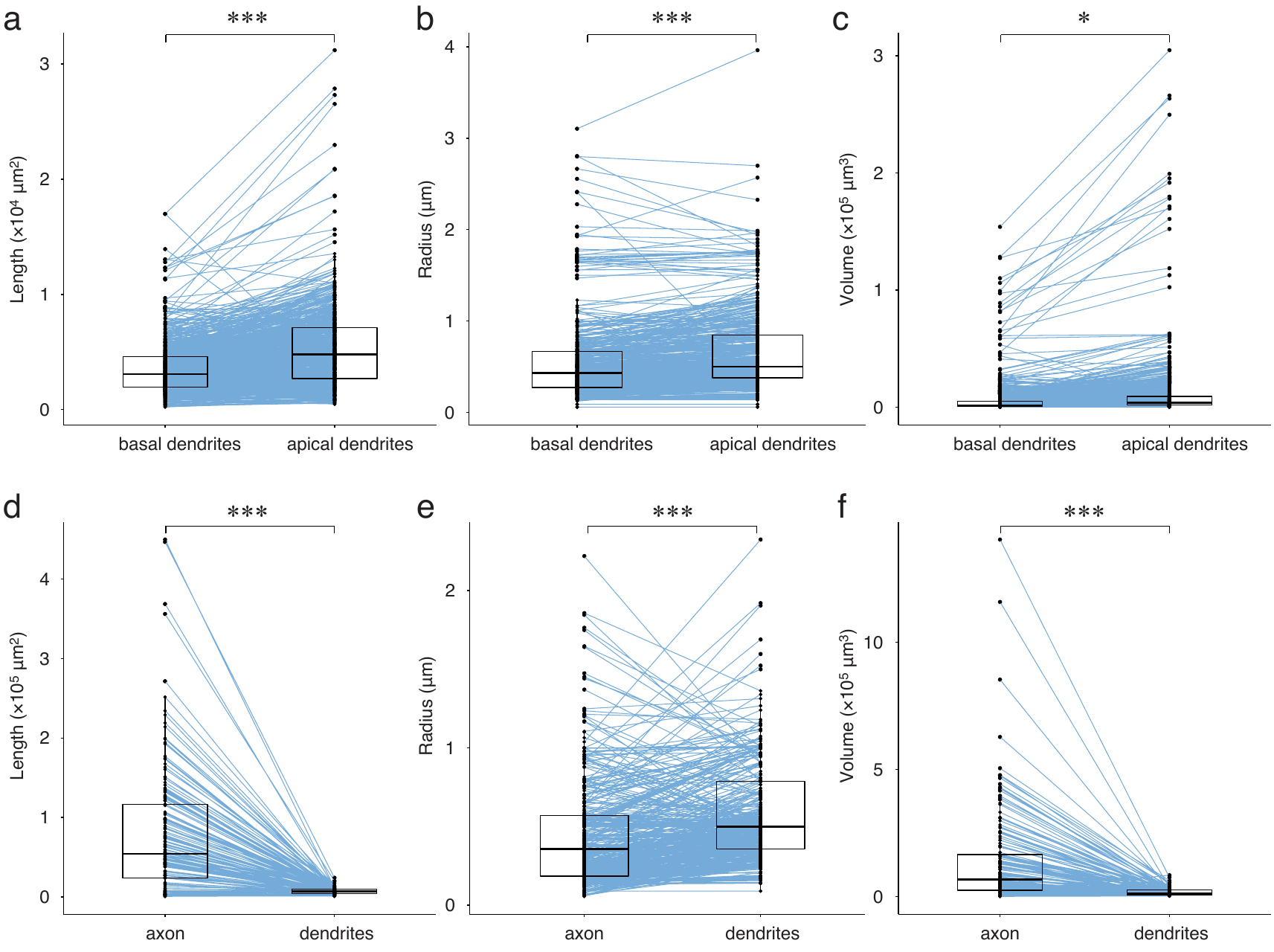}
\par\end{centering}

\caption{\label{fig:2}Paired box plots for morphometric measures in cortical pyramidal neurons.
Comparison of total length~$L$, average radius~$r$, and total volume~$V=\pi r^2 L$ was performed for basal dendrites versus apical dendrites (a-c) or axon versus dendrites (d-f).
Individual measurements are represented with black dots. Paired measurements performed in the same cell are connected with thin blue lines.
The bottom and the top of each box represent the lower ($Q1$) and upper ($Q3$) quartile, whereas the black line in the middle of the box represents the median.
The interquartile range (IQR = $Q3 - Q1$) contains the middle 50~\% of the data, the whiskers extending from the minimum $Q1 - 1.5\times$IQR to the maximum $Q3 + 1.5\times$IQR value indicate the spread of the data, and the outliers are represented by data points that are located outside the whiskers of the box plot.
Statistical significance was estimated by repeated-measures analysis of variance (rANOVA): *, $p < 0.05$; ***, $p<0.001$.}
\end{figure}

Pyramidal neurons are located within layers 2, 3, 5 and 6 of the neocortex \citep{Shipp2007}.
The cell body (soma) of pyramidal neurons has the shape of a pyramid
with its base facing towards the deeper layers and its apex towards
the superficial layers of the cerebral cortex \citep{Bekkers2011}.
Because the dendrites of pyramidal neurons from layers 2, 3 and 5 reach layer 1, the size and complexity of their dendritic trees increases with the depth of the neuron within the cortex. In contrast, the dendrites of layer~6 neurons reach only layer~4, which explain why their dendritic trees are smaller and less complex than layer 5 neurons (Figure~\ref{fig:1}).
On average across all types of cortical pyramidal neurons, the basal dendrites have $\approx33.6\%$ shorter total length ($3513\pm2199$~$\mu$m) in comparison to apical dendrites ($5295\pm3524$~$\mu$m) ($F_{1,737}=73.0$, $P<0.001$, Figure~\ref{fig:2}a).
The mean radius of basal dendrites is also $\approx14.0\%$ thinner ($0.54\pm0.40$~$\mu$m) compared
to apical dendrites ($0.63\pm0.40$~$\mu$m) ($F_{1,737}=30.8$, $P<0.001$, Figure~\ref{fig:2}b),
which results in $\approx47.6\%$ lower total volume of basal dendrites
($5881\pm14897$~$\mu$m\textsuperscript{3}) as opposed to apical
dendrites ($11231\pm29236$~$\mu$m\textsuperscript{3}) ($F_{1,737}=5.7$, $P=0.017$, Figure~\ref{fig:2}c).
Detailed morphometric data for basal and apical dendrites
in neocortex, subiculum or hippocampus of mouse, rat and human are
presented in Table~\ref{tab:1} and Table~\ref{tab:2}.

\pagebreak

\begin{table}[t]
\caption{\label{tab:1}Morphometric measures for basal dendrites of pyramidal
neurons.}
\centering
\begin{tabular}{|p{1.5cm}|p{2.3cm}|p{1.9cm}|p{1.4cm}|p{2.5cm}|p{2.5cm}|p{2cm}|}
\hline
Species & Brain region & Neuron type & \# of cells & Total length ($\mu$m) & Total volume ($\mu$m\textsuperscript{3}) & Mean radius ($\mu$m)\tabularnewline
\hline
\hline
Mouse & Neocortex & Layer 2-3 & 15 & $3398\pm1638$ & $8138\pm6050$ & $0.84\pm0.23$\tabularnewline
\cline{3-7}
 &  & Layer 5 & 156 & $3433\pm2439$ & $5702\pm8551$ & $0.64\pm0.40$\tabularnewline
\cline{3-7}
 &  & Layer 6 & 37 & $3728\pm1773$ & $8058\pm6309$ & $0.79\pm0.24$\tabularnewline
\cline{2-7}
 & Hippocampus & CA1 & 20 & $2342\pm755$ & $3568\pm5430$ & $0.59\pm0.27$\tabularnewline
\cline{2-7}
 & Subiculum & Principal & 24 & $3192\pm1081$ & $4179\pm3305$ & $0.62\pm0.20$\tabularnewline
\hline
Rat & Neocortex & Layer 2-3 & 125 & $2086\pm1120$ & $850\pm1122$ & $0.32\pm0.15$\tabularnewline
\cline{3-7}
 &  & Layer 5 & 171 & $4854\pm2416$ & $11473\pm26126$ & $0.60\pm0.39$\tabularnewline
\cline{3-7}
 &  & Layer 6 & 20 & $1966\pm445$ & $531\pm290$ & $0.28\pm0.06$\tabularnewline
\cline{2-7}
 & Hippocampus & CA1 & 113 & $2955\pm1364$ & $5255\pm14778$ & $0.50\pm0.64$\tabularnewline
\cline{3-7}
 &  & CA3 & 62 & $4358\pm2166$ & $3036\pm4122$ & $0.44\pm0.23$\tabularnewline
\hline
Human & Neocortex & Layer 2-3 & 6 & $8220\pm1153$ & $10453\pm2079$ & $0.64\pm0.04$\tabularnewline
\hline
\end{tabular}
\end{table}

\begin{table}[t!]
\caption{\label{tab:2}Morphometric measures for apical dendrites of pyramidal
neurons.}
\centering
\begin{tabular}{|p{1.5cm}|p{2.3cm}|p{1.9cm}|p{1.4cm}|p{2.5cm}|p{2.5cm}|p{2cm}|}
\hline
Species & Brain region & Neuron type & \# of cells & Total length ($\mu$m) & Total volume ($\mu$m\textsuperscript{3}) & Mean radius ($\mu$m)\tabularnewline
\hline
\hline
Mouse & Neocortex & Layer 2-3 & 15 & $3470\pm1735$ & $8463\pm6775$ & $0.83\pm0.24$\tabularnewline
\cline{3-7}
 &  & Layer 5 & 156 & $4067\pm2503$ & $6512\pm7853$ & $0.69\pm0.37$\tabularnewline
\cline{3-7}
 &  & Layer 6 & 37 & $3622\pm1715$ & $7691\pm5640$ & $0.79\pm0.24$\tabularnewline
\cline{2-7}
 & Hippocampus & CA1 & 20 & $3821\pm1119$ & $6658\pm8380$ & $0.67\pm0.26$\tabularnewline
\cline{2-7}
 & Subiculum & Principal & 24 & $3130\pm1591$ & $4276\pm4812$ & $0.61\pm0.21$\tabularnewline
\hline
Rat & Neocortex & Layer 2-3 & 125 & $2418\pm1192$ & $1471\pm2028$ & $0.39\pm0.20$\tabularnewline
\cline{3-7}
 &  & Layer 5 & 171 & $8393\pm4638$ & $28280\pm52456$ & $0.78\pm0.45$\tabularnewline
\cline{3-7}
 &  & Layer 6 & 20 & $3947\pm1360$ & $1767\pm786$ & $0.38\pm0.11$\tabularnewline
\cline{2-7}
 & Hippocampus & CA1 & 113 & $6517\pm1903$ & $10939\pm27586$ & $0.57\pm0.57$\tabularnewline
\cline{3-7}
 &  & CA3 & 62 & $6228\pm1940$ & $6170\pm5512$ & $0.54\pm0.24$\tabularnewline
\hline
Human & Neocortex & Layer 2-3 & 6 & $9163\pm1588$ & $12488\pm4609$ & $0.65\pm0.07$\tabularnewline
\hline
\end{tabular}
\end{table}

\begin{table}[t!]
\caption{\label{tab:3}Morphometric measures for axons of pyramidal neurons.}
\centering
\begin{tabular}{|p{1.5cm}|p{2.3cm}|p{1.9cm}|p{1.4cm}|p{2.5cm}|p{2.5cm}|p{2cm}|}
\hline
Species & Brain region & Neuron type & \# of cells & Total length ($\mu$m) & Total volume ($\mu$m\textsuperscript{3}) & Mean radius ($\mu$m)\tabularnewline
\hline
\hline
Mouse & Neocortex & Layer 2-3 & 15 & $65749\pm49489$ & $95778\pm41755$ & $0.75\pm0.23$\tabularnewline
\cline{3-7}
 &  & Layer 5 & 81 & $87865\pm 103291$ & $165888\pm 250718$ & $0.72\pm0.23$\tabularnewline
\cline{3-7}
 &  & Layer 6 & 37 & $86886\pm51082$ & $131152\pm107960$ & $0.67\pm0.22$\tabularnewline
\cline{2-7}
 & Hippocampus & CA1 & 1 & $24677$ & $20412$ & $0.51$\tabularnewline
\cline{2-7}
 & Subiculum & Principal & 24 & $47600\pm 25061$ & $81450\pm 98365$ & $0.64\pm0.20$\tabularnewline
\hline
\end{tabular}
\end{table}

\subsection{Axon morphometry}

Axons are specialized to deliver electric output to distant targets.
The mean radius of axons ($0.45\pm0.34$~$\mu$m) is $\approx25\%$
thinner compared to the mean radius of dendrites ($0.60\pm0.39$~$\mu$m)
(paired $t$-test, $t_{1,527}=10.3$, $P<0.001$, Figure~\ref{fig:2}e) estimated in a subset of 528 cells, which had either partial axonal arborizations in slices ($n=370$) or complete axonal arborizations in whole brain reconstructions ($n=158$). The rationale for this analysis is that slicing does not affect the radii of neuronal projections.
However, because the axonal arborizations are trimmed in slice sections, for the evaluation of total axonal length and total axonal volume, we have used only the subset of 158 automated whole brain reconstructions in mouse (Table~\ref{tab:3}).
Axons of mouse projection neurons have $\approx10.8\times$ greater total length ($79020\pm81159$~$\mu$m)
and $\approx8.1\times$ greater total volume ($137351\pm193481$~$\mu$m\textsuperscript{3})
in comparison with dendrite total length ($7319\pm4079$~$\mu$m)
(paired $t$-test, $t_{1,157}=11.5$, $P<0.001$, Figure~\ref{fig:2}d) and total volume
($16862\pm14656$~$\mu$m\textsuperscript{3}) (paired $t$-test, $t_{1,157}=8.1$, $P<0.001$, Figure~\ref{fig:2}f).

\subsection{Energetics of the cerebral cortex}

Approximately 20\% of resting oxygen consumption (i.e. in the absence of heavy physical work by skeletal muscles) is absorbed by the human brain \citep{Laughlin1998}.
Brain activity is fueled almost exclusively by glucose \citep{Magistretti2015}.
Oxidative metabolism in mitochondria of 1~glucose molecule leads to
the production of 32~ATP molecules \citep{Mergenthaler2013}, each
of which releases $0.4$~eV of free energy upon hydrolysis \citep{George1970,Scott2005}.
Thus, the free energy available for utilization by neuronal activities
from glucose is only $1235$~kJ/mol, even though combustion of glucose
in oxygen releases 2801~kJ/mol. From the speeds of glucose consumption
\citep{Herculano-Houzel2011} by the cerebral cortex of different species
(Table~\ref{tab:4}), it can be estimated that the power of the mouse
cortex is $0.004$~W, rat cortex is $0.015$~W, and human cortex is
$4.427$~W. These cortical values comprise approximately half of the
power of the whole brain (Table~\ref{tab:5}), namely, the power of
the mouse brain is $0.008$~W, rat brain is $0.025$~W, and human
brain is $9.628$~W. This modest consumption of energy points to highly
efficient energy utilization, and miniaturization of the brain's logical circuitry.

\begin{table}[t!]
\caption{\label{tab:4}Energy consumption by neurons in the cerebral cortex.}
\centering
\begin{tabular}{|p{1.5cm}|p{2.5cm}|p{2.5cm}|p{2.5cm}|p{2.5cm}|p{2.5cm}|}
\hline
Species & Cortical mass (gray + white matter) (g) & Glucose use per gram per minute ($\mu$mol/g$\cdot$min) & Total number of cortical neurons ($\times10^{7}$) & Energy use per neuron (pW) & Total energy use by cortex (W)\tabularnewline
\hline
\hline
Mouse & $0.173$ & $1.10$ & $1.369$ & $286.13$ & $0.004$\tabularnewline
\hline
Rat & $0.769$ & $0.95$ & $3.102$ & $484.77$ & $0.015$\tabularnewline
\hline
Human & $632.52$ & $0.34$ & $1634$ & $270.91$ & $4.427$\tabularnewline
\hline
\end{tabular}
\end{table}

\begin{table}[t!]
\caption{\label{tab:5}Energy consumption by neurons in the brain.}
\centering
\begin{tabular}{|p{1.5cm}|p{2.5cm}|p{2.5cm}|p{2.5cm}|p{2.5cm}|p{2.5cm}|}
\hline
Species & Brain mass (g) & Glucose use per gram per minute ($\mu$mol/g$\cdot$min) & Total number of brain neurons ($\times10^{7}$) & Energy use per neuron (pW) & Total energy use by brain (W)\tabularnewline
\hline
\hline
Mouse & $0.416$ & $0.89$ & $7.089$ & $107.50$ & $0.008$\tabularnewline
\hline
Rat & $1.802$ & $0.68$ & $20.013$ & $126.03$ & $0.025$\tabularnewline
\hline
Human & $1508.91$ & $0.31$ & $8606$ & $111.88$ & $9.628$\tabularnewline
\hline
\end{tabular}
\end{table}

\subsection{Computational capacity of pyramidal neurons}

\begin{figure}[t!]
\begin{centering}
\includegraphics[width=165mm]{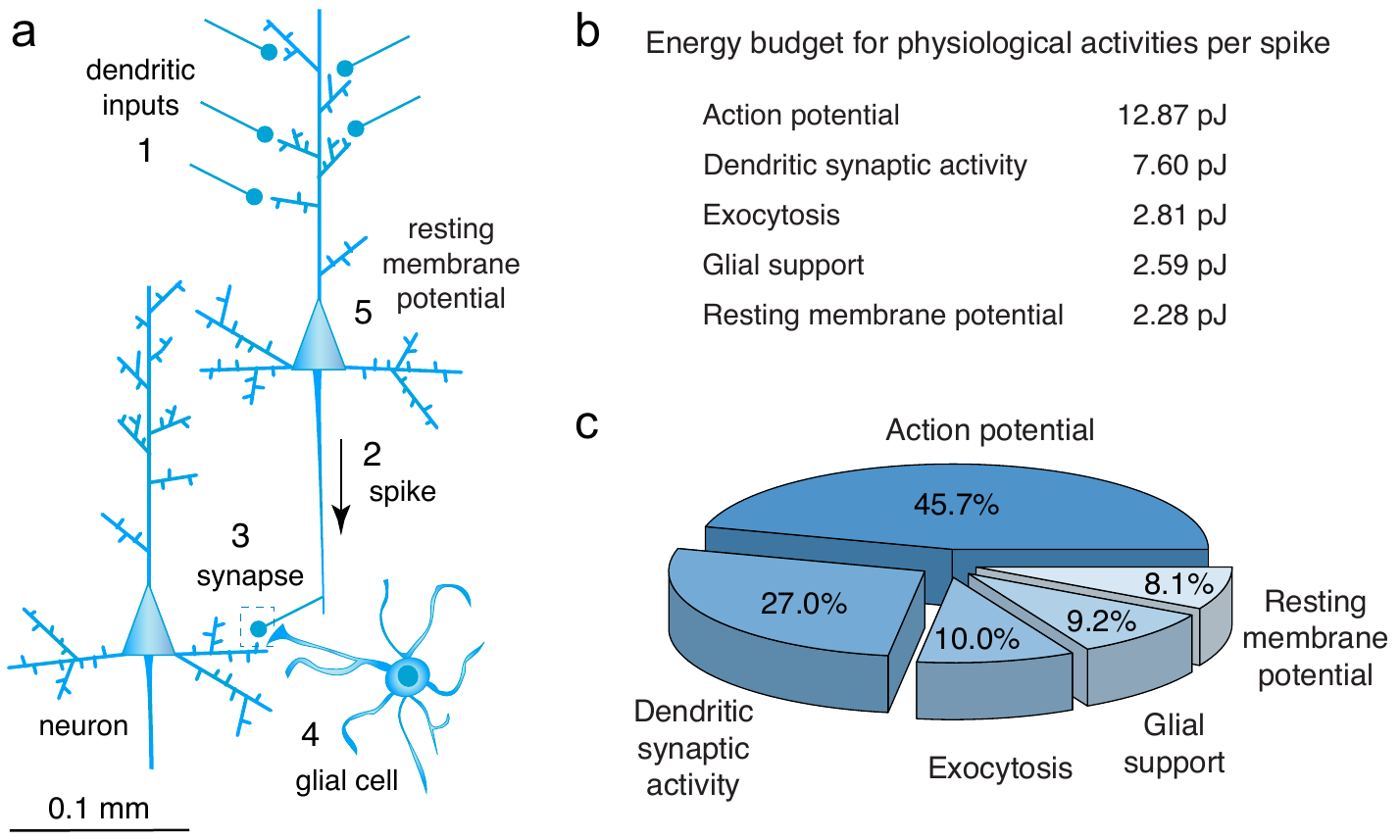}
\par\end{centering}

\caption{\label{fig:3}Physiological activities underlying the input, processing and output of cognitive information through electric spikes by pyramidal neurons.
(a) To generate an electric spike, each neuron (1) needs multiple excitatory dendritic inputs, which activate post-synaptic neurotransmitter receptors.
The excitatory post-synaptic potentials (EPSPs) then (2) summate at the soma and trigger an action potential at the axonal hillock.
The action potential propagates to pre-synaptic axonal buttons that (3) release neurotransmitter through exocytosis of synaptic vesicles.
Excess neurotransmitter is (4) recycled by glial cells, which support the proper functioning of neurons.
In between spikes, pyramidal neurons expend energy in order to (5) maintain their resting membrane potential.
The energy budget in picojoules (pJ) for these five main physiological activities per spike is tabulated in (b) and displayed as a pie chart with percentages of the total energy consumed in (c).}
\end{figure}

Pyramidal neurons input, process and output cognitive information with the use of electric spikes.
There are five main physiological processes that support each spike (Figure~\ref{fig:3}a):

(1) Each neuron needs multiple excitatory dendritic inputs, which activate post-synaptic neurotransmitter receptors and generate excitatory post-synaptic potentials (EPSPs).

(2) EPSPs propagate towards the soma where they summate to reach a certain voltage threshold and trigger an action potential at the axonal hillock.
The action potential then propagates to pre-synaptic axonal buttons that innervate target neurons.

(3) To excite the target neurons, the pre-synaptic electric spike is temporarily converted into chemical signal through exocytosis of synaptic vesicles that release excitatory neurotransmitter such as glutamate or aspartate.

(4) The excess neurotransmitter is recycled by glial cells. Glial cells also support neuronal homeostasis by controlling the chemical contents of the extracellular matrix.

(5) In between spikes, both pyramidal neurons and glial cells expend energy in order to maintain
their resting membrane potentials and to restore the initial concentration gradients of Na\textsuperscript{+}, K\textsuperscript{+} or Ca\textsuperscript{2+} ions.

The energy expenditure \emph{in vivo} varies from neuron to neuron depending on the exact morphometric measures and physiological activities.
However, it is possible to estimate the energy budget for an average cortical pyramidal neuron under the assumption that all of the energy released from glucose consumed by the brain cortex is used to fuel electric spiking with the underlying biomolecular processes.
Because the maintenance of resting membrane potential by neurons, and the accompanying glial support could be viewed as continuous processes interspersed by discrete action potentials, the general outline of the calculation of the energy budget per spike is as follows:
Firstly, from the total energy budget of the cerebral cortex, the energy needed for neuronal resting membrane potential and glial support is subtracted.
Secondly, from the remaining energy, the maximal average frequency of firing electric spikes is computed.
Thirdly, the maximal average firing frequency provides an estimate of the duration of the average interspike interval and the corresponding energy expenditure for the neuronal resting membrane potential and glial support per single spike.
Finally, the data so obtained will be integrated for calculating the total energy expenditure to support a single spike together with its preceding interspike interval.

\subsubsection{Resting membrane potential}

Physiological electric activities are due to passage of metal ions
across the plasma membrane. The resting membrane potential of neurons
is approximately $-70$ mV. The influx of Na\textsuperscript{+} ions leads
to depolarization, whereas the efflux of K\textsuperscript{+} ions
leads to hyperpolarization of the transmembrane voltage.
Neurons use $3.42\times10^{8}$ ATP molecules per second in order to keep steady
their resting membrane potential \citep{Attwell2001}. The power consumed
by $1.634\times 10^{10}$ neurons in the human cerebral cortex \citep{Herculano-Houzel2011} for their resting potential
is $0.358$ W, which accounts for $\approx8.1\%$ of the total cortical power.

\subsubsection{Glial support}

Glial cells also spend energy to sustain their resting membrane potential
at about $-60$ mV \citep{McKhann1997}. Glial cells, which are $3.8\times$
more numerous than neurons in the cerebral cortex \citep{Azevedo2009},
consume $1.02\times10^{8}$ ATP molecules per glial cell each second
\citep{Attwell2001}. For the human cerebral cortex, the energy
consumption by glial cells is $0.406$ W, which constitutes $\approx9.2\%$ of the
total cortical power.

\subsubsection{Action potentials}

Neuronal dendrites are unmyelinated and leak-prone, but they need to be
depolarized in their proximal part that is adjacent to soma, to a level slightly
above a threshold of $-54$ mV \citep{Pathak2016} in order to trigger an action potential at the axonal hillock. To reduce leakage,
and achieve efficient transport of the electric spike to distant targets,
the axons are myelinated with the exception of the nodes of Ranvier, where upon
electric stimulation, the membrane readily depolarizes to $+40$ mV
due to opening of voltage-gated Na\textsuperscript{+} channels. The
number of Na\textsuperscript{+} ions entering into a cylindrical
neurite segment is given by the capacitor charge formula
\begin{equation}
N=\frac{2\pi r L f \Delta V C_{m}}{q_{e}} = \frac{A f \Delta V C_{m}}{q_{e}} \label{eq:1}
\end{equation}
where $r$ is the radius,
$L$ is the length,
$A = 2\pi r L$ is the surface area,
$f$ is the fraction of unmyelinated active membrane,
$\Delta V$ is the voltage change,
$C_{m}=1$~$\mu$F/cm\textsuperscript{2} is the specific membrane capacitance,
and $q_{e}=160.218$~zC is the elementary electric charge.

Dendrites are completely unmyelinated $f=1$. They are depolarized
by $\Delta V=50$ mV during the backpropagation of an action potential
\citep{Attwell2001}. Direct substitution in Eq.~\eqref{eq:1} establishes that, for each action potential, in the basal dendrites
with mean total length $L=3513$~$\mu$m and average radius $r=0.54$
$\mu$m enter $N_\textrm{basal}=3.72\times10^{7}$ Na\textsuperscript{+} ions, whereas
in apical dendrites with mean total length $L=5295$~$\mu$m and average
radius $r=0.63$~$\mu$m enter $N_\textrm{apical}=6.54\times10^{7}$ Na\textsuperscript{+}
ions.

Axons are heavily myelinated with $f=0.018$ as estimated from the
mean length of unmyelinated nodes of Ranvier, which is $1.5$~$\mu$m,
and the internode mean distance of $81.7$~$\mu$m \citep{Arancibia-Carcamo2017}.
During an action potential, axons are depolarized by $\Delta V=110$
mV \citep{Schwindt1997}. Again, direct substitution in Eq.~\eqref{eq:1} establishes that, for axonal trees with mean total length $L=79020$
$\mu$m and average radius $r=0.45$~$\mu$m, enter $N_\textrm{axon}=2.76\times10^{7}$
Na\textsuperscript{+} ions per action potential.

For a pyramidal soma with surface area $A=2970\pm514$~$\mu$m\textsuperscript{2}
\citep{Zhu2000}, $f=1$ and $\Delta V=110$ mV, direct substitution in Eq.~\eqref{eq:1} establishes that there is an additional
load of $N_\textrm{soma}=2.04\times10^{7}$ Na\textsuperscript{+} ions.

For a realistic estimate of the total Na\textsuperscript{+} entry into a pyramidal
neuron per action potential, the amount of Na\textsuperscript{+}
ions computed from the capacitor charge formula should be multiplied by an overlap factor $f_\textrm{overlap}$
in order to account for simultaneous activation of Na\textsuperscript{+}
and K\textsuperscript{+} channels with Hodgkin--Huxley kinetics
\begin{equation}
N_\textrm{HH} = f_\textrm{overlap} \times \left(N_\textrm{basal} + N_\textrm{apical} + N_\textrm{axon} + N_\textrm{soma}\right) \label{eq:2}
\end{equation}
Computational simulations by \citet{Attwell2001} have found that $f_\textrm{overlap} = 4$.
The total sodium load in dendrites, soma and axon obtained from Eq.~\eqref{eq:2} amounts to
$N_\textrm{HH} = 6.024\times10^{8}$ Na\textsuperscript{+} ions. These Na\textsuperscript{+}
ions need to be pumped out of the neuron by protein Na\textsuperscript{+}/K\textsuperscript{+}-ATPase,
which exports 3 Na\textsuperscript{+} ions and imports 2 K\textsuperscript{+}
ions for every ATP molecule that is consumed \citep{SenguptaBiswa2013}.
Thus, for each electric spike, to remove the load of Na\textsuperscript{+} ions each neuron needs $N_\textrm{HH} /3 = 2.008\times10^{8}$ ATP molecules, which amounts to 12.87~pJ of energy (Figure~\ref{fig:3}b).

\subsubsection{Exocytosis of synaptic vesicles}

Exocytosis with subsequent endocytosis of a single synaptic vesicle consumes $1.24\times10^{4}$
ATP molecules \citep{Attwell2001}. Recycling of $4000$ glutamate
neurotransmitter molecules released per vesicle, through glutamate
uptake by glial cells, glial conversion to glutamine, export of glutamine
to neurons, neuronal conversion to glutamate and re-packaging into
synaptic vesicles, consumes another $1.1\times 10^4$ ATP molecules
\citep{Attwell2001}. Thus, the total energy consumption by a single
synaptic vesicle is $2.34\times 10^4$ ATP molecules.

Axons of cortical pyramidal neurons form between $7000$ and $8000$
synapses onto target neurons \citep{Braitenberg1998}. The number of released synaptic vesicles $N_\textrm{released}$ is proportional to the total number of axonal synapses $N_\textrm{synapses}$ and the probability of release $p_\textrm{release}$ of a synaptic vesicle per action potential per synapse
\begin{equation}
N_\textrm{released} = p_\textrm{release} \times N_\textrm{synapses}
\end{equation}
Considering that the release probability of a synaptic vesicle is only $0.25$
per action potential per synapse \citep{GeorgievGlazebrook2018}, each
pyramidal neuron will release on average $1875$ vesicles (for projection neurons most of the targets can be extracortical).
Thus, for each electric spike, the exocytosis of synaptic vesicles
consumes $4.39\times10^{7}$ ATP molecules, which amounts to 2.81~pJ of energy (Figure~\ref{fig:3}b).

\subsubsection{Synaptic activity in dendrites}

Action potentials cannot be spontaneously generated by healthy pyramidal neurons from their resting state.
Instead, a significant amount of preceding dendritic activity would be required to excite the neuron.
An important factor in many related scenarios is the nature and effect of spatial compartmentalization. For instance, \citet{Polsky2004} observed that rat neocortical pyramidal neurons initially process their synaptic inputs within thin dendritic subunits regulated by a nonlinear sigmoidal-type threshold, and then at a second stage they are linearly combined to deliver the overall neuronal response.

Each excitatory synaptic input delivered at a dendrite spine-head depolarizes the soma by only $0.12$~mV \citep{Kubota2015}. Thus, for a potential rise of 16~mV, from the resting membrane potential of $-70$~mV to the spike threshold of $-54$~mV, summation of at least 134 dendritic spine inputs would be needed. Indeed, based on detailed experimental electrophysiological data \emph{in vivo} using two-photon activation of an intracellular caged NMDA receptor antagonist, it was confirmed that the dendrites of pyramidal neurons need to receive an excess of excitatory synaptic inputs $N_\textrm{excess}=140$ (activating NMDA and AMPA receptors) in order to trigger an action potential \citep{Palmer2014}. 

In the awake state characterized by $\gamma$-frequency electric oscillations, the activation of powerful
perisomatic inhibition by fast-spiking interneurons \citep{Hu2014,Georgiev2014},
however, causes hyperpolarization or shunting that suppresses the
effects of excitatory synaptic activation in pyramidal neurons.
To take into account the effect of cortical inhibition in the presence of non-zero excitatory to inhibitory (E/I) ratio, the number of dendritic synaptic inputs $N_\textrm{inputs}$ can be modeled as
\begin{equation}
N_\textrm{inputs} = N_\textrm{excess} \times \left( 1 + \frac{1}{\textrm{E/I~ratio}} \right) \label{eq:4}
\end{equation}
The E/I ratio at the soma of layer 2/3 pyramidal neurons is $0.8$, whereas at the soma of layer 5 pyramidal neurons
it is $0.2$ \citep{Yang2018}.
Direct substitution in Eq.~\eqref{eq:4} shows that, in the presence of active inhibitory interneurons, the generation of an electric spike would require 315 dendritic inputs for layer 2/3 pyramidal neurons and 840 dendritic inputs for layer 5 pyramidal neurons.
For the estimation of the energy budget, we will take the average requirement of $578$ excitatory
synaptic inputs delivered to the dendritic tree for triggering of an action potential in a single cortical pyramidal neuron.

The release of a single synaptic vesicle filled with glutamate leads
to activation of post-synaptic NMDA and AMPA receptors whose opening
lets $3.8\times10^{5}$ Na\textsuperscript{+} ions and $10^{4}$
Ca\textsuperscript{2+} ions enter into the dendrite \citep{Attwell2001}.
Calcium signaling in dendrites leads to Ca\textsuperscript{2+} load
that needs to be removed by Na\textsuperscript{+}/Ca\textsuperscript{2+}
exchanger, which exports 1 Ca\textsuperscript{2+} ion and imports
3 Na\textsuperscript{+} ions. The 3 Na\textsuperscript{+} ions are
subsequently exported by Na\textsuperscript{+}/K\textsuperscript{+}-ATPase
consuming 1 ATP molecule. For $578$ synaptic inputs per action potential,
the extrusion of the Na\textsuperscript{+} and Ca\textsuperscript{2+}
ion load requires $7.9\times10^{7}$ ATP molecules. Recycling of the
vesicles (discussed in the preceding subsection) further requires $1.35\times10^{7}$ ATP molecules.
The energy expenditure for Ca\textsuperscript{2+} signaling during backpropagation
of the action potential from axonal hillock to dendrites adds another
$2.61\times10^{7}$ ATP molecules \citep{Attwell2001}.
Thus, for each electric spike, dendritic signaling consumes $1.186\times10^{8}$ ATP molecules in total, which amounts to 7.60~pJ of energy (Figure~\ref{fig:3}b).

\subsubsection{Rationing of the energy budget across physiological activities}

Taking stock of matters so far, we see that neural information is indeed costly. 
The energy expenses by a single neuron to support the dendritic synaptic activity required to elicit an action potential, to sustain the propagation of the action potential towards pre-synaptic axonal buttons, and to execute the associated release of synaptic vesicles for neurotransmitter
signaling, sum up to $3.633\times10^{8}$ ATP molecules, which release $23.28$ pJ of free energy.
For $1.634\times10^{10}$ neurons in the human cerebral cortex (Table~\ref{tab:4}), the required energy to fire once
is $0.38$~J. After subtraction from the total cortical budget of
the energies spent on the resting membrane potential by neurons and
glial cells, there is a remaining energy power of $3.663$~W that can
be spent by the cerebral cortex on firing action potentials with an
average frequency of $9.6$~Hz. This constitutes $\approx14.3\%$
of the maximal firing frequency of $67$ Hz that can be attained by
layer 5 pyramidal neurons \citep{Schwindt1997}.
For average spiking frequency of $9.6$~Hz, the duration of the interspike interval is $\approx104$~ms.
Thus, for each electric spike, the maintenance of neuronal resting membrane potential in the preceding interspike interval uses 2.28~pJ and the glial support uses 2.59~pJ of energy (Figure~\ref{fig:3}b).

The total energy budget for a single electric spike together with the preceding interspike interval amounts to 28.15~pJ.
In summary, 45.7~\% of the energy budget is dedicated for propagation of the action potential,
27.0~\% for support of dendritic synaptic activity,
10.0~\% for exocytosis of synaptic vesicles,
9.2~\% for glial support, and
8.1~\% for maintenance of the neuronal resting membrane potential in the interspike interval (Figure~\ref{fig:3}c).

Original estimates by \citet{Attwell2001} pointed to $3.29\times10^{9}$ ATP molecules (210.85~pJ) consumed by a neuron with a mean firing rate of 4~Hz. The energy budget stipulated 47\% for the production of action potentials, 34\% for the activity of dendritic post-synaptic receptors, 6\% for presynaptic exocytosis including recycling of excess glutamate, and 13\% for maintenance of the resting state of neurons and supporting glial cells. 
The main difference between that previous study and our present results stems from the precise morphometric data that we have used resulting in higher average spiking frequency by pyramidal neurons due to lower energy needs to support action potentials.

\subsubsection{The Landauer limit}

Moving a single elementary electric charge (electron, proton or monovalent
ion) across the plasma membrane through a potential difference of
$110$ mV dissipates $0.11$~eV of energy. The energy of $23.28$~pJ consumed
per action potential is sufficient for the motion of $1.32\times10^{9}$
elementary electric charges. The transport of an elementary electric
charge across the plasma membrane, however, may not be the elementary
bit of neuronal logical operation. Landauer's limit asserts that the
minimum possible amount of energy required by thermodynamics to erase one bit of information
(e.g. through application of an irreversible gate such as AND gate
or OR gate) is
\begin{equation}
E_{\min}=k_{B}T\ln2
\end{equation}
recalling that $k_{B}=1.38\times10^{-23}$ J/K
is Boltzmann's constant and $T$ is the absolute temperature \citep{Landauer1961}.
Otherwise expressed, if $\Delta E_{\rm{env}}$ denotes energy dissipated into the environment, and $\Delta S_{\rm{sys}}$ the thermodynamic entropy equivalent to information erased from the system memory, then
\begin{equation}
\Delta E_{\rm{env}} \geq  T \Delta S_{\rm{sys}}
\end{equation}
The total number of erased bits of information $I_\textrm{erased}$ from the system is bounded by \citep{Street2020,Bormashenko2019}
\begin{equation}
I_\textrm{erased} \leq \frac{\Delta E_{\rm{env}}}{k_B T \ln 2}
\end{equation}

\begin{figure}[t!]
\begin{centering}
\includegraphics[width=165mm]{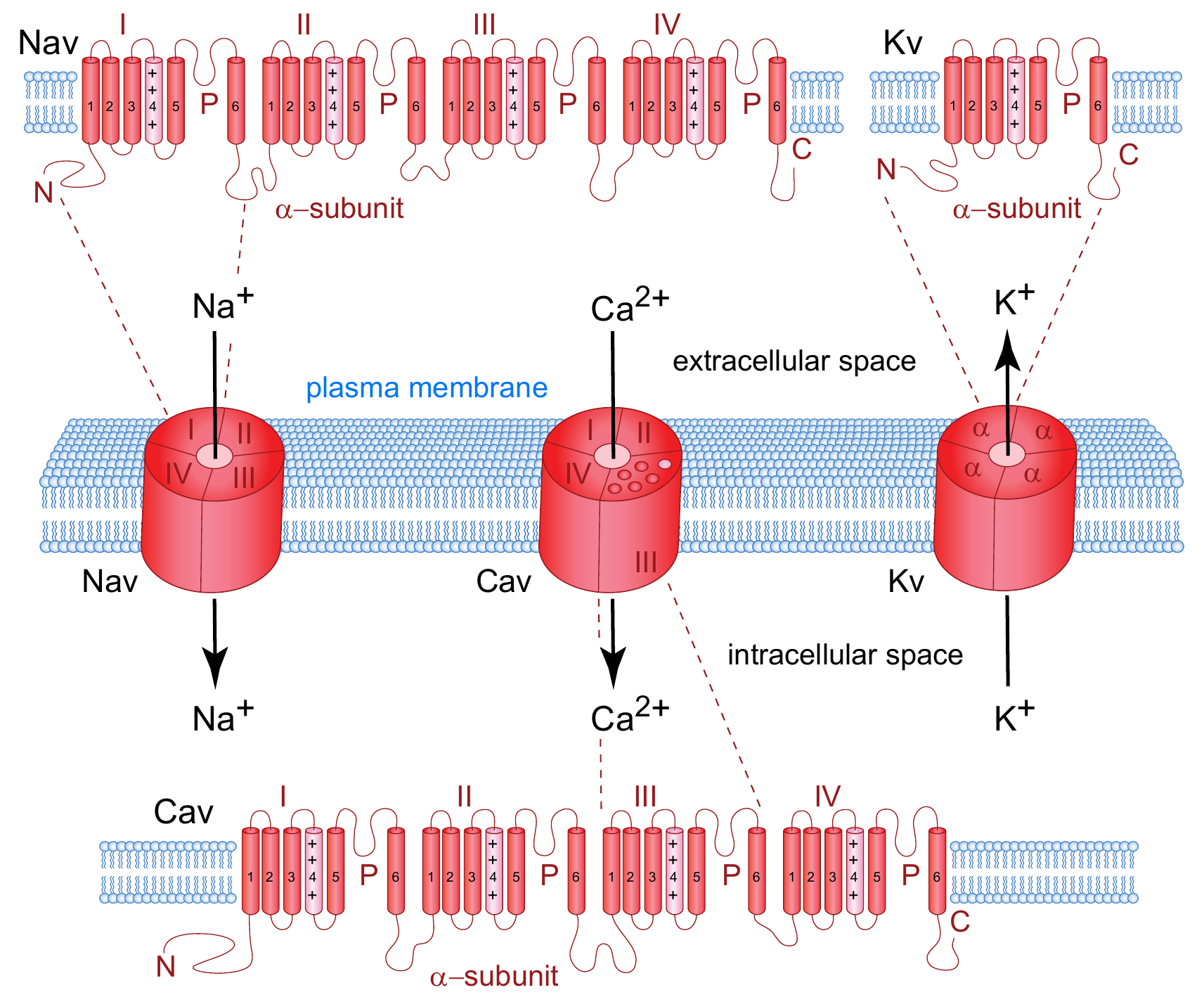}
\par\end{centering}

\caption{\label{fig:4}Electric activities of pyramidal neurons are generated
by sodium (Nav), potassium (Kv) and calcium (Cav) voltage-gated ion
channels incorporated into the plasma membrane, which consists of
a phospholipid bilayer with thickness of 10~nanometers. Structurally,
individual voltage-gated ion channels contain four protein domains
I-IV. Each domain has six transmembrane $\alpha$-helices (1-6). The
pore of the ion channel is configured by protein loops (P) connecting
the 5th and 6th $\alpha$-helices. Voltage sensing is accomplished
by the 4th $\alpha$-helix (S4), which usually carries six positively
charged lysine or arginine amino acid residues. Each proton tunneling event between
neighboring positively charged sites in the S4 voltage-sensor is ideally suited to represent a single Landauer elementary logical operation.}
\end{figure}

At physiological temperature of $310$ K, Landauer's limit is $2.968$ zJ ($18.526$ meV).
Therefore, the energy of $23.28$~pJ consumed per action potential is sufficient for the execution of $7.844\times 10^9$ Landauer elementary logical operations.
Noteworthy, the passage of a single elementary electric charge across the plasma membrane is equivalent
to $\approx5.94$ such elementary logical operations.
Since each S4 protein $\alpha$-helix voltage-sensor in voltage-gated
ion channels (Figure~\ref{fig:4}) usually contains 6 positively charged
amino acid residues \citep{Catterall1988}, the proton tunneling between
neighboring positively charged sites in the S4~voltage-sensor \citep{Kariev2007,Kariev2012,Kariev2018,Kariev2019}
is ideally suited to represent a single Landauer elementary logical
operation in cortical neural networks. Protons interact with water
and biological matter, mainly in a non-classical manner including
exchange-correlation effects, chemical bonding in hydronium-like complexes,
and tunneling \citep{Lobaugh1996}. Once the S4~protein $\alpha$-helix
voltage-sensors adopt an open channel conformation, the subsequent
flow of metal ions across the ion channel leads to amplification of
individual events of proton quantum tunneling that
occurred in the S4 voltage sensors \citep{Kariev2007,Kariev2012,Kariev2018,Kariev2019}.
This is how nanoscale quantum events may be amplified to exert macroscopic
effect on neuronal behavior and brain function \citep{Georgiev2013,Georgiev2020}.
If the energy power of $3.663$~W available for electric spiking is
completely miniaturized at the Landauer limit of $2.968$ zJ,
the human brain cortex will be able to execute the equivalent of over 1.2 zetta elementary logical
operations per second.

\section{Discussion}

In this work, we have evaluated the cognitive computational
capacity of the brain based on its experimentally measured glucose
consumption \citep{Herculano-Houzel2011}. The tightness of the bound is justified by two biomedical
facts. Firstly, the brain does not have an internal store of glucose,
but needs to rely on blood glucose level maintained by physiologically
regulated release from the liver glycogen depot \citep{Guyton2006}.
Secondly, the brain does not have anything like a long-term energy battery capable
of supporting cognitive computation in the absence of glucose, because
loss of clinical consciousness (syncope) occurs within seconds of
a sudden drop in blood glucose levels, or a brief cessation of cerebral
blood flow \citep{Kapoor2000}. This implies that the rate of glucose
consumption indeed puts a tight upper bound on the brain's capacity for cognitive
computation.

In the awake state, the energy power of $4.427$~W consumed by the human cerebral
cortex permits a maximal average spiking rate of $9.6$~Hz. This is
consistent with the observed average spontaneous firing rate of $12.87$~Hz by neurons in the visual cortex of awake rhesus monkeys \citep{Chen2009},
albeit it is somewhat higher than the average spontaneous firing rates
of $3.28$~Hz in rat \citep{Aasebo2017} or $1.88$~Hz in mouse \citep{Durand2016}.
Evoked activity of neurons in primary visual cortex of awake mice
in response to an optimal drifting grating, however, exhibited average
firing rates that were strongly dependent on locomotion: $2.9$~Hz
for stationary mice and $8.2$~Hz for mice running on a freely rotating
spherical treadmill with their heads fixed \citep{Niell2010}. Thus,
the estimated maximal average spiking rate of $9.6$~Hz might be reached
by an active brain engaged in a complex problem-solving task that
integrates multi-modal information from different senses.

Detailed analysis of neuronal morphology combined with energetics
of electrophysiological and molecular processes support the highly efficient
miniaturization of logical computational gates performed by pyramidal
neurons in the cerebral cortex. The energy utilized for the generation
of a single action potential is sufficient for the execution of $7.844\times10^{9}$
elementary logical operations. The cerebral cortex appears to have
attained the maximal computational efficiency allowed by Landauer's
thermodynamic limit: quantum tunneling of a proton between neighboring
positively charged S4~sensor sites in voltage-gated ion channels constitutes
a single Landauer elementary logical operation, whereas the transport
of a monovalent metal ion through the open ion channel pore constitutes
six such operations.

Landauer's limit sets ultimate energy constrains on the functioning of physical computing devices in the presence of a thermal bath \citep{Sagawa2008,Sagawa2010,Hong2016,Gaudenzi2018,Lent2019}. The original formulation put forward by \cite{Landauer1961} is motivated by consideration of the finite capacity of working memory of computing devices, namely the act of resetting of the working memory to its initial empty state requires compression of the phase space of the memory device, which will decrease its entropy. The second law of thermodynamics, however, requires that the total entropy of the memory device and its environment increases in time \citep{Leff2002}. Therefore, resetting the working memory must be accompanied by a corresponding entropy increase in the environment, in the form of heat dissipation, which is at least $k_{B}T\ln2$ joules per bit \citep{Landauer1961}.
The brain cortex, which is responsible for the stream of consciousness, allows us to store cognitive information only for short time periods before we forget it, or replace it with new sensory information. Thus, one interpretation of Landauer's principle is that the working of the human mind spends energy to forget \citep{Plenio2001}, where the energy dissipation occurs in the act of irreversible resetting of the cortical working memory.
An alternative formulation based on the theory of dissipative quantum channels, however, establishes that communication of classical information across a noisy quantum channel \citep{Jagadish2018} that is immersed in a heat bath with effective temperature $T$, also requires energy expenditure of at least $k_{B}T\ln2$ joules per bit \citep{Porod1984,Levitin1998}. Otherwise, the signal transmitted from the sender gate to the receiver gate could not be distinguished from the ambient thermal noise \citep{Porod1984,Levitin1998}. Thus, another interpretation of Landauer's principle is that the working of the human mind spends energy to transmit information between different noisy neuronal compartments (dendrites, soma, axon) or to communicate unambiguously with effector organs (e.g.~muscles through intermediate extracortical centers such as $\alpha$-motor neurons in the spinal cord).
It is likely that, in the course of an electric spike, cortical pyramidal neurons spend energy both for resetting their S4 voltage sensors in the resting ion channel state and for transmission of the electric signal from dendrites toward the axon terminals. Much related are implications of this neurophysiology together with Landauer's principle for human cognition, as discussed in \citet{Collel2015,Street2016,Street2020} with significant pointers towards variational free energy and its role in perceptual Bayesian inference \citep{Friston2010,Friston2013}.

Because physical dynamics at the nanoscale is able to manifest characteristic quantum mechanical effects, our results provide a rigorous foundation,
as far as energy considerations are concerned, for future development
of quantum models of the transmembrane electromagnetic field and its
interaction with mobile electric charges inside protein voltage-gated
ion channels \citep{Georgiev2017,Kariev2007,Kariev2012,Kariev2018,Kariev2019}
or membrane-bound SNARE proteins whose zipping mechanism triggers neurotransmitter
release \citep{GeorgievGlazebrook2018,GeorgievGlazebrook2019a,GeorgievGlazebrook2019b}.
Technological advances in available supercomputers have already led
to routine simulation of quantum dynamics of small biomolecules in
electrolyte solution with the use of quantum chemistry software implementing
density functional theory \citep{Kolev2011,Kolev2013,Kolev2018}. Applications
of recent theorems in quantum information, as based on generalized
uncertainty relations \citep{Carmi2019} to quantum brain states \citep{Georgiev2013,Georgiev2020}
may further shed light on the perplexing open problems in the cognitive sciences.

To summarize, we have implemented fundamental physical principles, including
the thermodynamically allowable Landauer's limit of energy spent on
elementary logical operations, to show that not all biomolecular processes
may contribute to cognitive computation, but mainly those involving
transmembrane proteins, such as voltage-gated or ligand-gated ion
channels, integrated into the electrically excitable neuronal plasma
membrane. Even though the human cerebral cortex may perform over 1.2~zetta logical operations per second,
exceeding over four orders of magnitude the capacity of modern supercomputers, 
we expect the implementation of large-scale and ultra in-depth
brain simulations to significantly advance in the foreseeable future.

\section{Conflict of Interest}

The authors certify that they have no affiliations with or involvement in any organization or entity with any financial interest, or non-financial interest in the subject matter or materials discussed in this work.

\section{Methods and Materials}

\subsection{Selection criteria for the morphometric study}

Morphometric parameters such as radii and lengths of neuronal projections constrain the electric performance of neurons and determine the number of physical charges that need to cross the plasma membrane in order to elicit a certain change in the transmembrane voltage.
For accurate assessment of the average radii and total lengths of different neurites (basal dendrites, apical dendrites, and axons), we have analyzed the full collection of pyramidal neuronal reconstructions in rodent (mouse, rat) or human brain cortex from NeuroMorpho.org~7.8 digital archive \citep{Ascoli2007} that pass the following selection criteria:
Firstly, we have selected only control experimental conditions with animals that did not express genetically-engineered disease-related protein mutations and were not exposed to pharmacological agents or harmful stimuli (e.g.~stress).
Secondly, only animals whose age corresponds to human age of over 3 months old were included. The utilized piecewise linear conversion formulas into corresponding human age are given for mice by \cite{Sengupta2013}, and for rats by \cite{Dutta2016}.
Thirdly, to ensure minimal trimming of dendritic trees for analysis of apical
and basal dendrites, we have included only reconstructions with minimal
slice thickness of 300~$\mu$m. Analysis of complete axonal arborizations
was performed in neuronal reconstructions from brain-wide imaging
data \citep{Economo2016,Gerfen2018}.
To verify the quality of all reconstructions, neurons were visualized in Neuromantic version~1.6.3 (\url{https://www.reading.ac.uk/neuromantic/body_index.php}) and .swc files with non-standard labeling of neurites or visually incomplete dendritic tree (e.g. apical dendrite was trimmed near its base) were excluded from further analysis.
Standardized .swc files are tables with 7 columns of numerical data for cable-like cylindrical segments that comprise the neuronal reconstruction (Table~\ref{tab:6}).
The lengths and volumes of neurite
segments was quantified with the use of custom Excel macros fetching
the cable radii and computing the Euclidean distances from the $x,y,z$ coordinates given in the .swc files. Morphometric data
are reported as mean $\pm$ standard deviation.

\begin{table}[h!]
\caption{\label{tab:6}Table structure of standardized .swc files.}
\centering
\begin{tabular}{|p{1.7cm}|p{2.9cm}|p{1.7cm}|p{1.7cm}|p{1.7cm}|p{1.5cm}|p{2.4cm}|}
\hline
1 & 2 & 3 & 4 & 5 & 6 & 7 \tabularnewline
\hline
\hline
segment number & structure identifier & $x$ position & $y$ position & $z$ position & radius $r$ & parent segment\tabularnewline
\hline
 integer value starting from~1 & 1 - soma\newline
2 - axon\newline
3 - basal dendrite\newline
4 - apical dendrite & coordinate in $\mu$m & coordinate in $\mu$m & coordinate in $\mu$m & segment radius in $\mu$m & parent segment number;\newline $-1$ is used for lack of parent\tabularnewline
\hline
\end{tabular}
\end{table}

\subsection{Neuronal reconstructions}

Digital reconstructions of pyramidal neurons in control experimental
conditions were selected from three animal species: mouse (252 neurons),
rat (491 neurons) and human (6 neurons). This dataset of 749 neurons
includes contributions from 32 labs: Amaral \citep{Ishizuka1995},
Arnold\_Johnston \citep{Arnold2019}, Barrionuevo \citep{Henze1996},
Blackman \citep{Blackman2014}, Buchs \citep{Larkum2004}, Chandrashekar
\citep{Economo2016}, Claiborne \citep{Carnevale1997}, De Koninck \citep{Bories2013},
Dendritica \citep{Vetter2001}, Feldmeyer \citep{Marx2012,Marx2015},
Groen \citep{Groen2014}, Hay \citep{Hay2013}, Helmstaedter \citep{Helmstaedter2008},
Hoffman \citep{Hoffmann2015}, Jaffe \citep{Chitwood1999}, Johnston
\citep{Dougherty2012,Malik2016}, Kawaguchi \citep{Hirai2012,Ueta2013},
Kole \citep{Hamada2015,Hamada2016,Hallermann2012,Kole2004,Kole2007,Kole2011},
Korngreen \citep{Bar-Yehuda2008}, Krieger \citep{Groh2009,Krieger2007},
Luo \citep{Gong2016}, Markram \citep{Anastassiou2015}, Martina \citep{Kelly2016},
MouseLight \citep{Gerfen2018}, Orion \citep{Santamaria-Pang2015},
Segev \citep{Eyal2016}, Soltesz \citep{Lee2014}, Spruston \citep{Golding2005},
Staiger \citep{Staiger2016}, Storm \citep{Honigsperger2015}, Topolnik
\citep{Tyan2014,Francavilla2018} and Urban \citep{Zhou2015,Tripathy2015}.

\subsection{Modeling of cortical layers}

Vector .svg images of individual neurons were rendered with HBP Neuron Morphology Viewer \citep{Bakker2016,Bakker2017} and scaling information was extracted with NeuroM, a Python-based toolkit for the analysis and processing of neuron morphologies developed by the
Blue Brain Project (\url{https://neurom.readthedocs.io/en/stable/}).
Modeling of the brain cortex in mouse was then performed in Adobe Illustrator based on measured thickness of cortical layers in Nissl stained coronal slices \citep{Franklin2007,Georgiev2016}.
All data from NeuroMorpho.Org digital archive was used in compliance with the online Terms of Use (\url{http://neuromorpho.org/useterm.jsp}).
In particular, all original papers that describe the reconstructions are cited, the complete name of the digital archive is clearly stated, attribution to the developers of the archive is given \citep{Ascoli2007}, and specific reconstructions are referenced with their NeuroMorpho.Org ID numbers.

\subsection{Statistical analysis}

Statistical analysis of neuronal morphology was performed using SPSS
ver.~23 (IBM Corporation, New York, USA). Comparison of morphometric
measures for apical and basal dendrites was performed with repeated-measures
analysis of variance (rANOVA) implemented as a general linear model
in which within-subject variable was dendrite type, between-subject
factors were animal species, brain region and neuronal type, and covariate
was slice thickness. Comparison of axons with dendrites was performed
with paired $t$-tests for a subset of the neuronal reconstructions
for which the axonal trees were complete.
Paired box plots were created with the use of ggpubr library in R~ver.~4.0.2 (R~Foundation for Statistical Computing, Vienna, Austria, \url{https://www.r-project.org/}).

\subsection{Energy consumption and energy units}

The energy consumption by pyramidal neurons in different animal species (mouse, rat or humans) was estimated
based on brain mass, glucose use per gram per minute, and total number of neurons in the brain or the cerebral cortex reported in \cite{Herculano-Houzel2011}.
For each molecule of glucose, oxidative metabolism in mitochondria produces 32~ATP molecules \citep{Mergenthaler2013}.
Hydrolysis of 1 ATP molecule releases $0.4$~eV of free energy \citep{George1970,Scott2005}, which is equivalent to 64.0872~zJ.
Thus, the useable energy from 1 glucose molecule is 12.8~eV (2050.79~zJ).
For molecular processes the energy consumption was reported in electron~volts~(eV), where 1~eV is the amount of energy required to move 1~electron across an electric potential difference of 1~V.
For macroscopic processes the energy consumption was reported in joules~(J), using a conversion formula where 1~eV equals to $160.218$~zJ.
Brain power was reported in watts~(W), where 1~W is defined to be the energy transfer at a rate of 1~J per second.
For all reported quantities standard SI~prefixes were used.

\section*{Acknowledgements}

E.C. acknowledges support from the Israel Innovation Authority under project 70002 and from the Quantum Science and Technology Program of the Israeli Council of Higher Education.

\end{document}